\documentclass[onecolumn,final]{svjour2}
\usepackage{amstext}
\usepackage{amssymb}
\usepackage{graphicx}
\usepackage{amsmath}
\usepackage{color}

\makeatletter
\@ifundefined{textcolor}{}
{%
 \definecolor{BLACK}{gray}{0}
 \definecolor{WHITE}{gray}{1}
 \definecolor{RED}{rgb}{1,0,0}
 \definecolor{GREEN}{rgb}{0,1,0}
 \definecolor{BLUE}{rgb}{0,0,1}
 \definecolor{CYAN}{cmyk}{1,0,0,0}
 \definecolor{MAGENTA}{cmyk}{0,1,0,0}
 \definecolor{YELLOW}{cmyk}{0,0,1,0}
 }

\makeatother

\usepackage[english]{babel}
\newcommand{\bx}{{\bf x}} 
\newcommand{\bsig}{\boldsymbol{\sigma}}

\newcommand{\cD}{{\cal D}}
\newcommand{\cS}{{\cal S}}
\newcommand{\cJ}{{\cal J}}
\newcommand{\cDo}{\cD^{(1)}}
\newcommand{\cDt}{\cD^{(2)}}

\newcommand{\Qt}{Q^{(2)}}

\begin{document}

\title{On the sufficiency of pairwise interactions in maximum entropy
  models of biological networks}

\author{Lina Merchan \and Ilya Nemenman} 

\institute{Lina Merchan\thanks{Present address: Department of
    Radiology, Emory University, Atlanta, GA 30322, USA} \at
  Department of Physics, Emory University, Atlanta, GA 30322, USA \and
  Ilya Nemenman \email{ilya.nemenman@emory.edu} \at Departments of
  Physics and Biology, Emory University, Atlanta, GA 30322, USA}
\date{\today}

\maketitle

\titlerunning{Pairwise interactions in maximum entropy models}

\begin{abstract}
  Biological information processing networks consist of many
  components, which are coupled by an even larger number of complex
  multivariate interactions. However, analyses of data sets from
  fields as diverse as neuroscience, molecular biology, and behavior
  have reported that observed statistics of states of some biological
  networks can be approximated well by maximum entropy models with
  only pairwise interactions among the components. Based on
  simulations of random Ising spin networks with $p$-spin ($p>2$)
  interactions, here we argue that this reduction in complexity can be
  thought of as a natural property of densely interacting networks in
  certain regimes, and not necessarily as a special property of living
  systems. By connecting our analysis to the theory of random
  constraint satisfaction problems, we suggest a reason for why some
  biological systems may operate in this regime.\end{abstract}

\keywords{collective dynamics\and $p$-spin models \and numerical
  simulations}

\section{Introduction}

The increased throughput of biological experiments now allows joint
measurements of activities of many basic components underlying
collective information processing in biological systems.  Such
multivariate data must be interpreted within models. Within this
context, Maximum Entropy (MaxEnt) models \cite{Jaynes:1957ua} have
been some of the most successful. The logic of such models is that,
ultimately, one wants to find an approximation $Q (\bx)$ to the joint
probability distribution $P(\bx)$ of the observed multivariate data
$\{x_i\}=\bx$, $i=1,\dots,N$. Unfortunately, for a large number of
components, $N$, the datasets can never be large enough to estimate
$P$ directly from data. One may only be able to estimate various
expectation values of functions of the data,
$\langle f_\kappa(\bx)\rangle_P=\tilde f_\kappa$,
$\kappa=1,\dots,K$. Then one can search for $Q$ that matches the
reliable estimates. If additionally one requests that $Q$ has no
structure beyond that required by the matching, then this is
equivalent to asking for $Q$ with the maximum entropy, subject to the
constraints imposed by the matching,
\begin{equation}
  Q=\arg\max S(Q)- \sum_\kappa\lambda_\kappa( \langle f_\kappa\rangle_Q- \tilde{f}_\kappa),
\end{equation}
where the entropy $S$ is defined as 
\begin{equation}
  S(Q)\equiv S(\bx)=-\sum_\bx Q(\bx)\log_2Q(\bx).
\end{equation}

A common special case of this general formulation is when the
variables are binary, which we will denote as
$x_i=\sigma_i\in\{-1,1\}$, and the data constrain their various
low-order correlation functions, such as $\langle \sigma_i\rangle$ or
$\langle \sigma_i\sigma_j\rangle$. In this case, the MaxEnt
approximation $Q$ is \cite{Schneidman:2003ty}:
\begin{equation}
  Q(\bsig)=\frac{1}{Z} \exp\left(-\sum_ih_i\sigma_i -
    \sum_{ij}J_{ij}\sigma_i\sigma_j
    -\sum_{ijk}K_{ijk}\sigma_i\sigma_j\sigma_k-\cdots\right).
\label{Qgeneral}
\end{equation}
Here every constrained correlation function gets a term in the
exponent, $Z$ is the partition function, and the Lagrange multipliers
$h_i, J_{ij},K_{ijk},\dots$ must be chosen to satisfy the constraints.
This is generally not an analytically solvable problem, and even
numerics are hard
\cite{Ackley:1985ce,Broderick:2007wq,Weigt:2009ba,Mezard:2009kx,Sessak:2009ei,Cocco:2011fo,Nguyen:2012wa}.

Equation~(\ref{Qgeneral}) has the form of the Ising spin problem,
allowing a wholesale import of intuition from statistical physics to
MaxEnt data analysis. Correspondingly, these ideas have been applied
to many biological systems in the last decade \cite{Bialek:2014wv},
starting with neurophysiological recordings from salamander retina
\cite{Schneidman:2006he}. There $N$ was a few dozen neurons, and
$\sigma_i=\pm1$ corresponded to the $i$'th neuron spiking/not spiking
at a certain time. A surprising result was that truncating
Eq.~(\ref{Qgeneral}) at the quadratic order in $\sigma_i$ (or, in
other words, constraining $Q$ up to pairwise correlations) provided a
good fit to $P$. We will refer to this finding as {\em pairwise
  sufficiency} from now on.

The pairwise sufficiency was later found in other neural systems
\cite{Tang:2008kg,Ohiorhenuan:2010bu,Field:2010gx} (though it is
violated at larger $N$ \cite{Tkacik:2014ey}). It was observed further
for natural images \cite{Bethge}; for discrete, yet non-binary $\bx$
in sequencing data \cite{Halabi:2009jc,Mora:2010jx}; and for
real-valued velocities of birds in flocking experiments
\cite{Bialek:2012kj}. Even for some non-MaxEnt approaches, similar
findings were also reported
\cite{Margolin:2010eo,Otwinowski:2013eq}. One can interprete these
observations in the context of biological systems operating in a
special regime \cite{Mora:2010hp,Schwab:2014io}. However, the wide
applicability of the findings suggests an alternative: pairwise
sufficiency may emerge for a wide class of biological and
non-biological networks {\em generically}. Indeed, sparse sampling of
variables in experiments is similar to decimation in statistical
physics, and the resulting renormalization group-like flow may
decrease the importance of the higher order couplings
\cite{Tkacik:2006vq}. Further, in a perturbative regime, where
fluctuations away from the independence are small, the pairwise
sufficiency also appears \cite{Roudi:2009eb}. Here we propose one more
possibility, arguing that the pairwise sufficiency arises naturally in
{\em strongly coupled} multivariate systems.

In what follows, we first introduce the idea in an intuitive toy
model, and then develop it numerically by analyzing randomly generated
networks. We show that the pairwise MaxEnt models approximate such
random networks surprisingly well. Further, we explore distributions
of states of these networks and their models, leading to an
explanation of the pairwise sufficiency.  Finally, we discuss why
diverse biological systems may find themselves in the
pairwise-sufficient regime, but we live it for the future to
investigate if this mechanism is responsible for the sufficiency in
experimental networks.

\section{Results}

\subsection{Building Intuition: Networks of {\tt XOR}s}
For a tractable example of emergence of the pairwise sufficiency, we
focus on Boolean gates. These are the limit of Ising spin networks in
the low temperature (strong coupling) regime
\cite{Schneidman:2003ty,Mezard:2003vv}. For example,
$\sigma_3=\sigma_1\, {\tt OR}\, \sigma_2$ can be written as
$P(\sigma_3|\sigma_2,\sigma_1)=\frac{1}{Z}\exp[J(\sigma_1\sigma_3+\sigma_2\sigma_3+\sigma_3)]$
with $J\to\infty$. If also $P(\sigma_1=\pm1)=P(\sigma_2\pm1)=1/2$,
then
$1/4P(\sigma_3|\sigma_2,\sigma_1)=P(\sigma_1,\sigma_2,\sigma_3)$. Thus
the joint probability distribution for {\tt OR} has the pairwise
MaxEnt form, Eq.~(\ref{Qgeneral}). Similarly, for $J\to\infty$,
$\sigma_3=\sigma_1\, {\tt AND}\, \sigma_2$ is equivalent to
$P(\sigma_1,\sigma_2,\sigma_3)=\frac{1}{Z}\exp[J(\sigma_1\sigma_3+\sigma_2\sigma_3-\sigma_3)]$. This
is again a pairwise MaxEnt distribution. However,
$\sigma_3=\sigma_1\, {\tt XOR}\, \sigma_2$, is equivalent to
$P(\sigma_1,\sigma_2,\sigma_3)=\frac{1}{Z}\exp(-K\sigma_1\sigma_2\sigma_3)$,
$K\to\infty$. This is an example of a purely third-order gate, with no
pairwise contributions to its MaxEnt representation.

\begin{figure}[t]
\centering
\includegraphics[height = 4in]{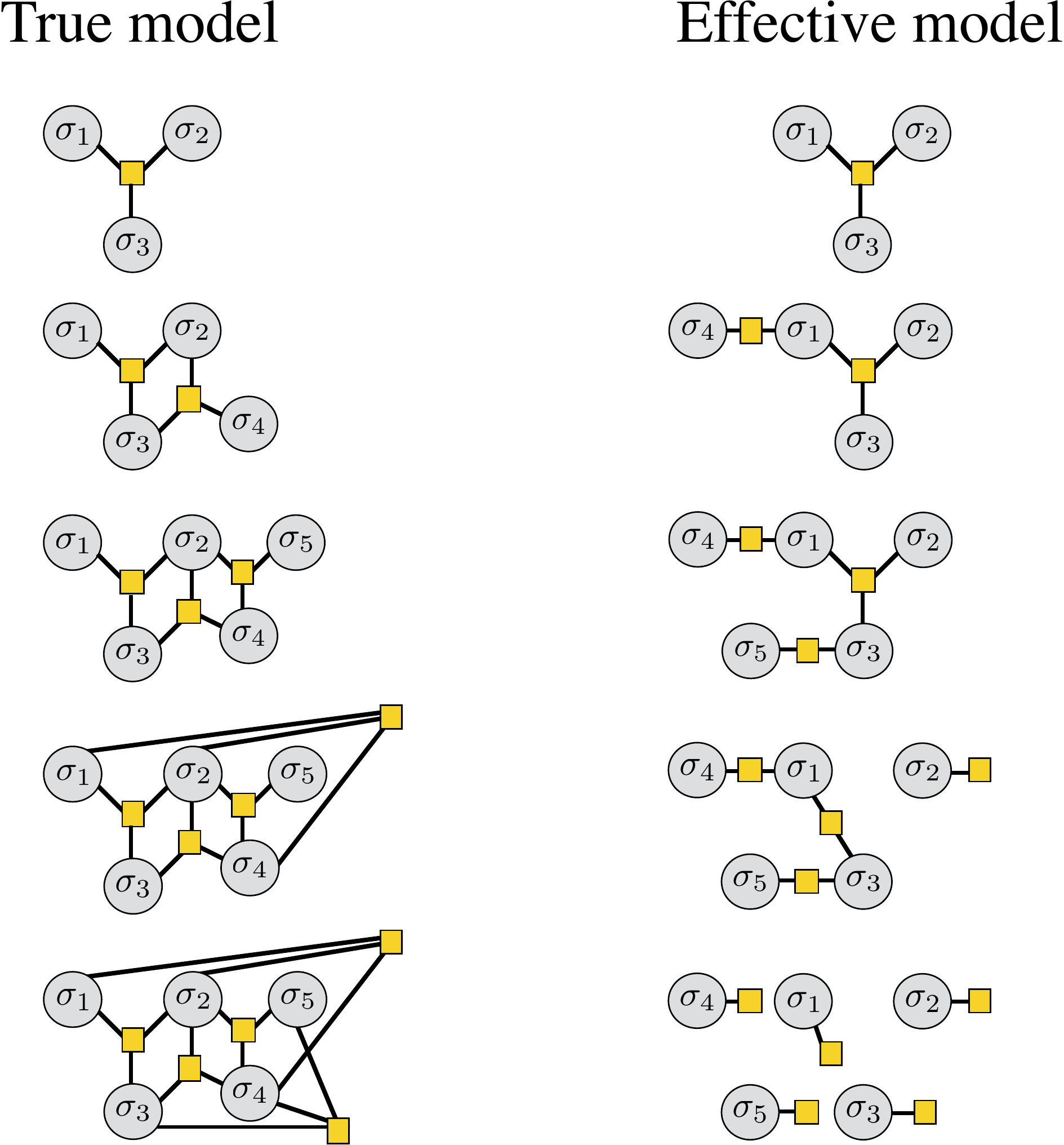}
\caption{\label{XORs}{\bf Emergence of pairwise interactions in a
    network of {\tt XOR} gates.} On the left, we show small networks
  of spins $\sigma_i$ (grey circles). The spins interact (yellow
  squares) by means of third order {\tt XOR} interactions. On the
  right, an equivalent network is shown, where some of the {\tt XOR}s
  get replaced by {\tt EQUAL} and assignment operations, which are the
  second and the first order interactions, respectively.}
\end{figure}

In Fig~\ref{XORs}, we now couple of a few such third-order gates to
each other. The spins $\sigma_1,\sigma_2,\sigma_3$ are connected by an
{\tt XOR} (left column, first row), and there is no simpler effective
representation of the network (right column, first row). We then add
the fourth spin, $\sigma_4=\sigma_2\, {\tt XOR}\, \sigma_3$ (left
column, second row). However, then $\sigma_4=\sigma_1$. This can be
represented as an effective model
$P(\sigma_4|\sigma_1,\dots,\sigma_3)=P(\sigma_4|\sigma_1)=\frac{1}{Z}\exp(J\sigma_1\sigma_4)$,
$J\to\infty$. Thus the third order {\tt XOR} interaction is equivalent
to a pairwise {\tt EQUAL} interaction (right column, second row). The
latter is effective and nonlocal, in the sense that $\sigma_4$ is
coupled to $\sigma_1$, with which it does not interact in the true
model. We further add $\sigma_5=\sigma_2\, {\tt XOR}\, \sigma_4$
(third row), and this is equivalent to an effective model
$\sigma_5=\sigma_3$. In short, of the three third order interactions,
each constraining one spin and hence ``carrying'' 1 bit of
information, two can be represented without any error as pairwise
interactions.  Now the network can exist in four distinct global
states out of $2^5=32$, determined by $\sigma_{1,2}=\pm1$ (namely,
$\bsig^{(1)}=\{-1,-1,-1,-1,-1\}$, $\bsig^{(2)}=\{-1,+1,+1,-1,+1\}$,
$\bsig^{(3)}=\{+1,-1,+1,+1,+1\}$, and
$\bsig^{(4)}=\{+1,+1,-1,+1,-1\}$). Thus it is far from the
perturbative regime of Ref.~\cite{Roudi:2009eb}.  We can grow the
network further so that each new spin is coupled by a third order
interaction to two existing spins. Then the number of spins, $N$, and
the number of interactions, $M$, are related as $N=M+2$, and all but
one third order interaction can be represented as a second order
one. In other words, an effective pairwise model has an error of only
$1/(N-2)$ when accounting for the statistics of the network states.

Alternatively, we can add more {\tt XOR}s without adding new
nodes. This may be inconsistent or redundant with already existing
couplings. Or in a case such as
$\sigma_1=\sigma_2\,{\tt XOR}\,\sigma_4$ (fourth row), this sets
$\sigma_2= -1$ (thus adding the bias, or the first order term), and
all other spins are equal to each other, so that the pairwise
effective model is exact. Finally, adding
$\sigma_3=\sigma_4\,{\tt XOR}\,\sigma_5$ sets every spin to -1, and
makes even the first-order model exact (bottom row).

We see that a network of {\tt XORs} can exhibit the pairwise sufficiency
nonperturbatively.  Of course, more realistic physical or biological
systems are stochastic ($J,K< \infty$), and such simple arguments will
not work. However, the example suggests that effective pairwise models
can approximate more complex networks well when nodes in the network
interact strongly and densely, and the space of network states is
sufficiently constrained. In such cases, there are many pairs of nodes
that are relatively strongly correlated simply by chance, allowing
replacement of higher order interactions with pairwise ones. In what
follows, we will develop this intuition further using numerical
studies.

\subsection{Pairwise approximations to random networks with higher
  order interactions}

To verify our intuition, we proceed by generating random networks that
have {\em only} higher order nondeterministic interactions among spins
($p$-spin models, $p>2$ \cite{Kirkpatrick:1987ve}). We then quantify
the accuracy with which lower order MaxEnt models approximate these
networks. We explore networks with $p=3,4$ to ensure that our findings
do not depend on the exact structure of the true higher order
interactions. Further, systems with only fourth order couplings have
the $Z_2$ symmetry, and thus cannot include any first order terms in
their MaxEnt approximations, Eq.~(\ref{Qgeneral}).  Studying them will
allow us to understand if the eventual freezing to a single
well-defined state, as in the last row of Fig.~\ref{XORs}, is crucial
for the pairwise sufficiency, or if it emerges even for
nonperturbative networks with more than one highly probable state.

To generate the random networks, we first specify $N$, the number of
nodes, and $M$ the number of interactions. Then for each interaction
$\mu=1,\dots,M$, we generate its coupling constant $K_\mu$ from a
zero-mean Gaussian distribution with a certain variance $s^2$. We then
choose three or four spins at random to couple. The overall
probability of states for these networks is 
\begin{align}
  P_3(\bsig)&=\frac{1}{Z}\exp\left(-\sum_{\mu=1}^M K_\mu
              \sigma_{\mu_1}\sigma_{\mu_2}\sigma_{\mu_3}\right),\;\mbox{3-spin model,}\label{triplet}\\
  P_4(\bsig)&=\frac{1}{Z}\exp\left(-\sum_{\mu=1}^M K_\mu
              \sigma_{\mu_1}\sigma_{\mu_2}\sigma_{\mu_3}\sigma_{\mu_3}\sigma_{\mu_4}\right),\;\mbox{4-spin
              model},\label{quartic}
\end{align}
where $\mu_1<\mu_2<\mu_3<\mu_4$, so that the spins do not
self-interact.  To specify these distributions (and later calculate
various errors of approximations), we need to know $Z$. To decouple
studying the problem of the pairwise sufficiency from a hard problem of
efficient sampling, we focus on $N\le 22$, which allows us to estimate
$Z$ by direct summation fast enough to do it many times and collect
statistics. We generate many such distributions $P_3$ and $P_4$, every
time picking random $N\in[10,22]$, $M\in [1,250]$, and
$s\in [0.2,2.0]$.

For each generated distribution, we estimate its individual and
pairwise marginals $P(\sigma_i)$, $P(\sigma_i,\sigma_j)$ for all
$i,j=1,\dots,N$ by direct marginalization (hereafter we drop
subscripts 3 or 4 for $P$ if it does not cause confusion).  We then
calculate the first order (or independent) MaxEnt approximation
\begin{equation}
Q^{(1)}(\bsig)=\prod_{i=1}^N P(\sigma_i).
\label{ind}
\end{equation}

Next we fit the pairwise MaxEnt model $Q^{(2)}$ to $P$.  While good
algorithms exist for this purpose
\cite{Ackley:1985ce,Broderick:2007wq,Weigt:2009ba,Mezard:2009kx,Sessak:2009ei,Cocco:2011fo,Nguyen:2012wa},
it is unclear if their assumptions are satisfied by our
networks. Trying again to decouple the problems of efficient inference
and the pairwise sufficiency, we choose a classic, well understood
Iterative Proportional Fitting Procedure (IPFP) algorithm
\cite{Deming:1940ug}. That is, we start with $Q^{(1)}$ as a guess for
$Q^{(2)}$, calculate $Q^{(2)}(\sigma_i,\sigma_j)$, and redefine
\begin{equation}
Q^{(2)}(\bsig)\to Q^{(2)}(\bsig)\frac{P(\sigma_i,\sigma_j)}{Q^{(2)}(\sigma_i,\sigma_j)}.
\end{equation}
We cycle through all pairs $i,j$, and iterate until
$Q^{(2)}(\sigma_i,\sigma_j)\approx P(\sigma_i,\sigma_j)$ up to a
relative error of $10^{-5}$. This is achieved within $\sim10^0\dots10^4$
iterations depending on how close the final $Q^{(2)}$ is to $Q^{(1)}$.
We verified that starting with different initial conditions results in
the same solution, as it should.

To measure the quality of the MaxEnt models, we calculate the
Kullback-Leibler (KL) divergence between the true distribution $P_{3,4}$ and each
approximation, normalized by the number of spins in the system:
\begin{eqnarray}
  {\cal D}^{(1)}=\frac{D_{\rm KL}^{(1)}}{N}&\equiv\frac{1}{N}\sum_{\bsig}
                                             P(\bsig)\log_2\frac{P(\bsig)}{Q^{(1)}(\bsig)},\\
  {\cal D}^{(2)}=\frac{D_{\rm KL}^{(2)}}{N}&\equiv\frac{1}{N}\sum_{\bsig}
                                             P(\bsig)\log_2\frac{P(\bsig)}{Q^{(2)}(\bsig)}.
\end{eqnarray}
Since our maximum $N$ is rather small, this is done by direct
summation. Notice that both $\cDo$ and $\cDt$ are between zero
(perfect fit) and one (the worst fit) if single-spin marginals of $Q$
and $P$ are equal.

\begin{figure}[t]
\centering
\noindent\includegraphics[height = 1.55in]{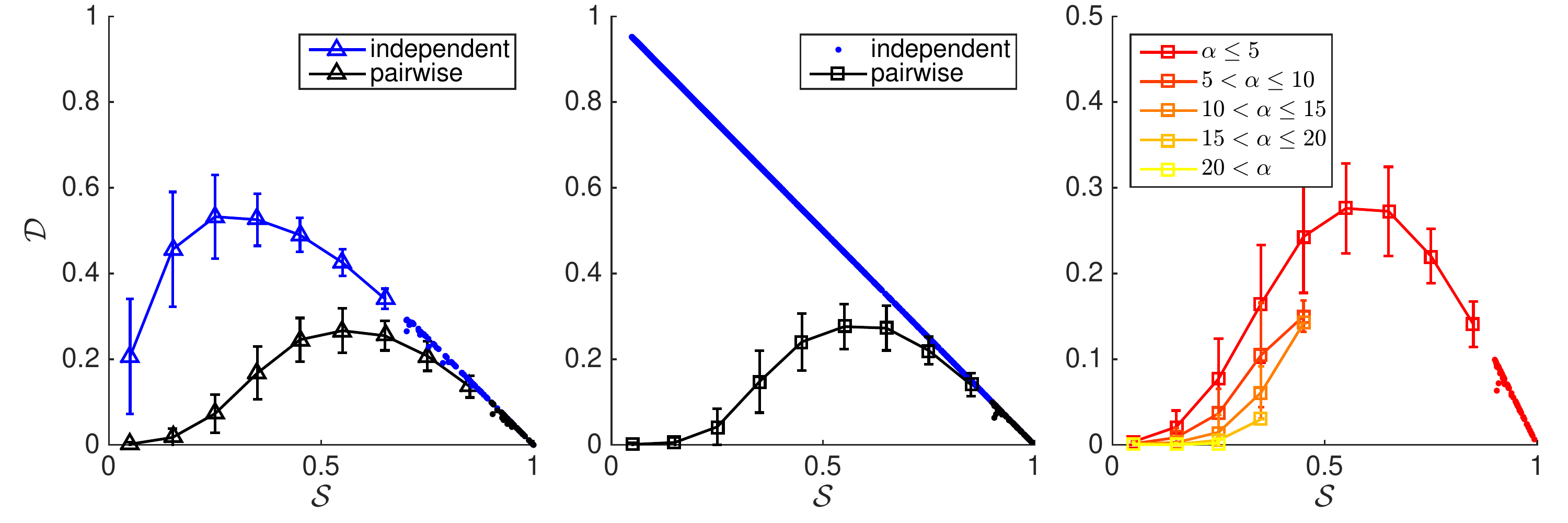}
\caption{\label{DvsS}{\bf Error of the MaxEnt fits vs.~the normalized
    entropy of the network state space, $\cS$.} The left panel shows
  errors of the independent, $\cDo$, and the pairwise, $\cDt$,
  approximations for 3-spin networks. We used $\sim800$ random
  networks with $11\le N\le20$ spins and with a varying number of
  interactions, $M$. We partitioned all the networks by their $\cS$ in
  bins of width of 0.1 and calculated the mean and the standard
  deviation of $\cD$ for each bin. These are indicated by triangles
  and the error bars. Wherever the data points for individual networks
  showed little scatter, we plotted these points instead of the bin
  averages. The middle panel presents similar data, $\cDo$ and $\cDt$,
  for 4-spin networks. Here over 4000 random networks were generated
  with $11\le N\le 22$.  $\cDt$ was again averaged within ten bins,
  and the means and the standard deviations are plotted. For $\cDo$,
  data for individual networks are presented.  These merge into a
  perfect straight line due to the $Z_2$ symmetry of 4-spin
  distributions. For both the 3- and the 4-spin cases, the pairwise
  sufficiency is clear at low $\cS$. The right panel replots the
  $\cDt$ data for the 4-spin networks, but splits them according to
  $\alpha$, which measures the average strength of interactions per
  spin within a network. Large $\alpha$ curves are significantly below
  their small $\alpha$ counterparts, indicating that, other things
  being equal, densely and strongly interacting networks are more
  likely to be pairwise sufficient. Notice that large $\alpha$ curves
  end abruptly since such networks cannot have large $\cS$. }
\end{figure}

In Fig.~\ref{DvsS}, we plot the values of ${\cal D}^{(1)}$ and
${\cal D}^{(2)}$ measured over different ensembles of random networks
vs.~the normalized entropy of the network's state space
$\cS=S(\bsig)/N$, which also varies between 0 and 1. For all types of
networks and approximation, the quality of fit is high
($\cD^{(\cdot)}$ is low) when $\cS\sim1$, so that the networks are
unconstrained, and nearly all states are possible. This is trivial
since even the zeroth order approximation (each spin up or down with
50\% probability) would work well here.

As $\cS$ decreases, the fit errors increase. When $\cS$ reaches small
values, the independent approximation, $\cDo$, starts behaving
differently for the different network types. In the 4-spin case, by
construction, $P(\bsig)=P(-\bsig)$. Thus $\langle\sigma_i\rangle=0$
for any $i$, and the best independent approximation is the uniform
distribution. For this construction, the smallest possible entropy is
$\cS=1/N$, where the network exists in two mirror states, and there
the error of $Q^{(1)}$ is $\cD_4^{(1)}=1-1/N$. In contrast, a 3-spin
network freezes at $\cS=0$, and each spin is strongly biased (as in
our {\tt XOR} networks above). Thus the independent approximation
provides a perfect fit in this case.

The distinction between $P_3$ and $P_4$ vanishes for the pairwise
MaxEnt approximation. Here, for both 3- and 4-spin networks, the fit
errors behave similarly: for $\cS$ decreasing from 1, $\cD^{(2)}$
grows from 0 and reaches its peak at about
$\cD^{(2)}\approx 0.25\dots0.3$ near $\cS\approx 0.5\dots0.6$. This is
already interesting: $\cDt$ almost never goes above 0.3 for all
networks we tried. Thus even in some of the worst cases, pairwise
approximation is quite good!  Further, for even smaller entropies,
$\cDt$ rapidly drops, approaching zero faster than linearly in
$\cS$. For $\cS\approx0.25$, $\cDt_3\approx0.07$. It is even smaller,
$\cDt_4\approx0.04$, for the quartic case. This is because
$\min \cS_4=1/N$, so that the whole $\cDt_4$ curve is slightly shifted
compared to $\cDt_3$ at low $\cS$. 

In summary, for all the networks we have considered, pairwise
sufficiency emerges robustly at low (but not too low) entropy. In
fact, at $\cS\approx0.25$, our networks can be in more than
$2^{S}=2^{N\cS}\approx2^5=32$ highly probable states. Thus the
networks are not totally frozen, and yet the pairwise approximation is
nearly sufficient! Crucially, this finding is robust to the changes in
the network size: $\cD$ vs.~$\cS$ curves are stable over the entire
range of $N$ we explored.

Within a single narrow bin of $\cS$, $\cDt$ may still have a rather
large range. We explore this variability in the rightmost panel of
Fig.~\ref{DvsS}. For this, we define $\alpha=sM/N$ (recall that $s$ is
the standard deviation of the random couplings used to generate the
networks). $\alpha$ measures the strength of interactions (or
constraints) per spin, analogously to a similar parameter in the
random constraint satisfaction problems
\cite{Mezard:2002ec,Krzakala:2007dd}.  For quartic networks, where we
have enough samples, we then plot $\cDt$ vs.~$\cS$ for different
ranges of $\alpha$. Crucially, we find that, for the same $\cS$, a
larger $\alpha$ results in better pairwise fits. In other words, a
denser and stronger interacting network is more likely to be pairwise
sufficient. This is potentially a good news for MaxEnt approaches to
biological systems, which are known for the immense complexity of the
underlying biophysical interactions.

\begin{figure}[t]
\centering
\includegraphics[height = 1.55in]{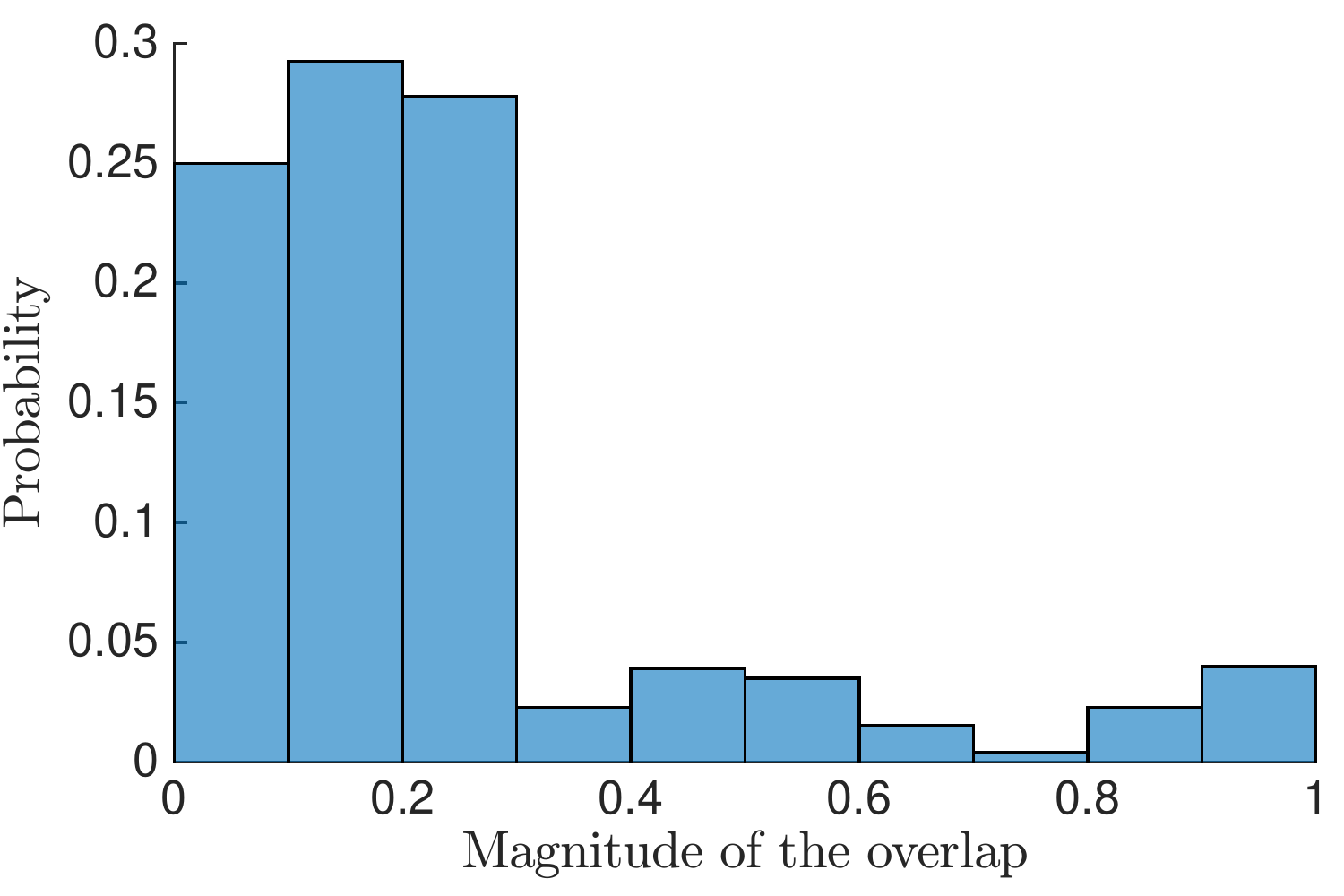}
\caption{\label{nonperturbative}{\bf The pairwise sufficiency emerges
    in the nonperturbative regime.}  We select all highly probable
  states $\bsig^\mu$, defined somewhat arbitrary as
  $P(\bsig^\mu)>0.001$. We then calculate the overlap for all pairs of
  such states. Finally, we plot the histogram of the magnitudes of the
  overlaps from all 4-spin networks with $N\ge20$, and with
  $0.1\le\cS\le0.3$. Such networks can exist in many states, but still
  very few compared to $2^N$. Most of the overlap magnitudes are away
  from 1, indicating small to moderate similarity among the highly
  probable states. Thus these states are broadly dispersed and do not
  cluster together.}
\end{figure}

We conclude this section by stressing that high probability states of
pairwise sufficient $p$-spin networks are not close to each other. To
illustrate this, we focus on the 4-spin case with $N\ge 20$, and on
small but not negligible $\cS$. We then evaluate the magnitude of the
overlap, $\left|\bsig^\mu\cdot\bsig^\nu\right|/N$, among all highly
probable network states and plot the distribution of the overlaps in
Fig.~\ref{nonperturbative}. For purely randomly distributed states, we
would expect the standard deviation of overlaps to be $\sim0.22$, and
a peak near zero. And we would expect magnitudes of overlaps near 1 if
all highly probable states were clustered near a dominant
one. Instead, the distribution in Fig.~\ref{nonperturbative} is not
concentrated near 1, and the standard deviation is
$\approx0.39$. Therefore, there is some clustering of probable states,
but certainly not strong clustering.  Thus the state space of our
networks cannot be described as small fluctuations around a dominant
state, and the pairwise sufficiency here is not perturbative
\cite{Roudi:2009eb}. It likely emerges due to a previously not
investigated mechanism.

\subsection{The structure of the state space of the pairwise
  sufficient networks}

The toy example of the {\tt XOR} network suggests that the pairwise
sufficiency may emerge when the network ``freezes'' to a few (but not
necessarily just one) highly probable states, and different relatively
tightly coupled clusters of spins decouple from each other. Is this
also true for our networks with a nonzero temperature? How do energy
landscapes of the sufficient and the insufficient networks differ from
each other? And are the MaxEnt fits for both cases structurally
different?

\begin{figure}[t]
\centering
\includegraphics[height = 1.7in]{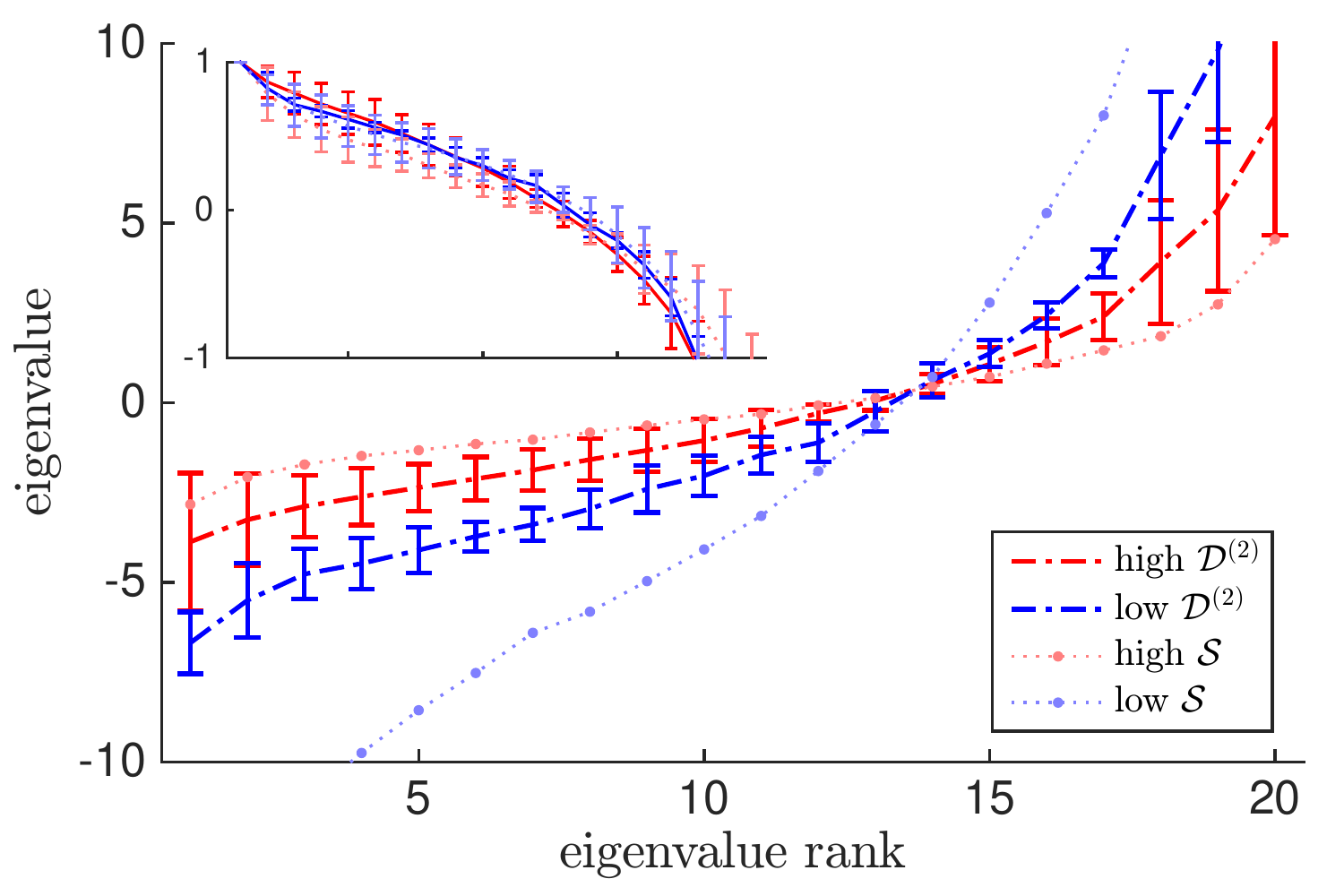}
\caption{\label{eigenvalues}{\bf Spectra of the pairwise coupling
    matrices $J_{ij}$ for MaxEnt approximations to random 4-spin
    networks, $N\ge20$.} We order eigenvalues from the smallest
  (lowest energy) to the largest (highest energy). We then plot the
  mean spectra (with standard deviations, where it does not obstruct
  the figures), averaged over different subsets of 4000 networks. Low
  $\cS$ subset corresponds to $0.1\le\cS<0.2$. Such networks are fit
  extremely well by pairwise models, with mean
  $\cDt\approx6\cdot10^{-3}$. High $\cS$ corresponds to
  $0.3\le\cS<0.4$. Here the pairwise fits are bad, so that the mean
  $\cDt\approx0.15$. Finally, for the intermediate range
  $0.2\le \cS<0.3$, the quality of fits is diverse. We further
  partition this range into well fitted, $\cDt\le0.06$, and badly
  fitted, $\cDt\ge 0.12$, subsets, leaving intermediate fits off the
  plot. The four average spectra show that the pairwise sufficiency is
  directly correlated with the scale of the spectra, with larger
  magnitude eigenvalues resulting in smaller $\cDt$. The inset shows
  the averages for each of the four ranges, where each spectrum is
  normalized by its largest magnitude negative eigenvalue. The four
  curves are very close for much of their range.  }
\end{figure}

To start exploring this, we estimate $h_i$ and $J_{ij}$ for $Q^{(2)}$
inferred using IPFP. We do this by choosing $\gg N(N+1)/2$ states with
the highest probability from $\Qt$. We get the energy of each such
state as $E(\bsig)=-\log P(\bsig)$ and then solve the linear
regression problem to find the coupling constants from the states and
their energies. Finally, we calculate the eigenvalue spectra of the
inferred $J_{ij}$, having set $J_{ii}=0$.  The averaged spectra are
shown in Fig.~\ref{eigenvalues} for different combinations of $\cS$
and $\cDt$. We see that the success of $Q^{(2)}$ is correlated with
the magnitude of the eigenvalues of $J_{ij}$ --- larger magnitudes,
which correspond to stronger interactions and more constrained
distributions, give the pairwise sufficiency. This is true
irrespective of $\cS$ (though $\cS$ and $\alpha$ are dependent, as we
have discussed). Crucially, if one rescales the spectra by their
largest magnitude negative eigenvalue (Fig.~\ref{eigenvalues}, inset),
then all spectra collapse. Thus the $J_{ij}$ (or the energy
landscapes) of the pairwise sufficient and the pairwise insufficient
fits are not intrinsically different: a rescaling (change in
temperature) can morph one into the other.

\begin{figure}[t]
\centering
\includegraphics[height = 1.5in]{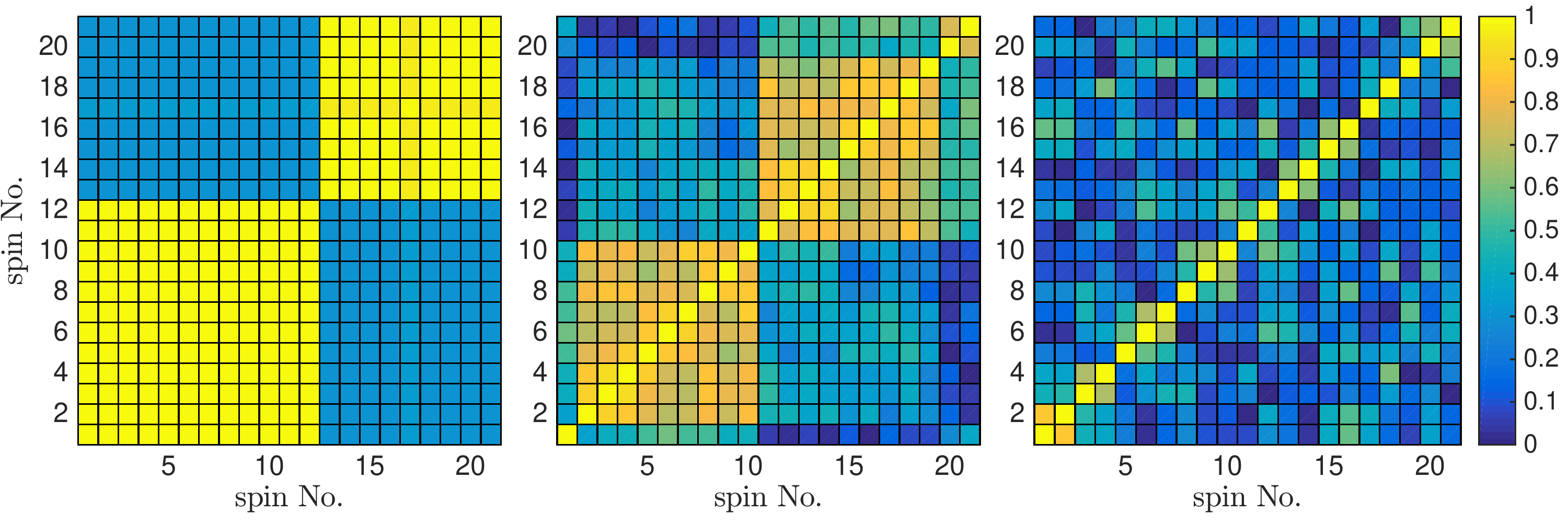}
\caption{\label{correlations}{\bf 4-spin networks decouple into spin
    clusters in the pairwise sufficient regime.} The three panels show
  typical values of $\left|c_{ij}\right|$ for $N=21$, sorted into
  correlated clusters. Left panel: a nearly perfectly pairwise
  sufficient network with $S=2.05$ bits and $\cDt=8.0\cdot10^{-5}$. In
  accord with the entropy value, the network splits into two clusters,
  with spins nearly perfectly correlated within each, but almost
  independent across. Middle panel: a good pairwise fit with $S=3.9$
  bits and $\cDt=0.032$. Correspondingly, four clusters of different
  sizes are seen. However, now the spins also exhibit some
  correlations across clusters, which presumably leads to the increase
  in $\cDt$. Right panel: a network with $S=5.7$ bits and
  $\cDt=0.16$ -- a bad (though not disastrous) pairwise fit. There are
  now many small clusters, but correlations within and across the
  clusters are not very different.}
\end{figure}

Having analyzed the pairwise sufficient and insufficient solutions,
$Q^{(2)}$, we now focus on the landscapes of the $p$-spin networks
themselves. The freezing that results in the decrease of $\cDt$ and
the growth of the eigenvalues of $J$ can create the landscapes of
different types. For example, the highly probable states may be
essentially uncorrelated, reminiscent of the landscapes of the
Hopfield network in the ferromagnetic phase
\cite{Hopfield:1982vm}. Alternatively, as in our {\tt XOR} network,
entire blocks of spins can merge into strongly correlated clusters,
which then decouple from each other. Then the low energy network
states will be direct products of the states of the clusters. To
disambiguate the two scenarios, we calculate pairwise spin-spin
correlation
$c_{ij}=\frac{{\rm cov}\, (\sigma_i,\sigma_j)}{{\rm std}\, \sigma_i
  {\rm std}\, \sigma_j}$
by direct summation of $P$ (note that, for 4-spin networks, the
correlation is equal to the covariance since
$\langle \sigma_i\rangle=0$). We then cluster the spins based on the
absolute value of their correlations.  Figure~\ref{correlations} shows
the clusters for 4-spin networks (note that since the number, the
size, and the spin assignment for clusters are different for each
network, we only show typical cases). A network with a near-zero
$\cDt$ (left panel) shows a perfect partitioning into two clusters;
$S(\bsig)\approx 2$ bits is a result of this partitioning. As networks
with larger entropies are considered, the number of clusters
increases, and their boundaries become fuzzy, leading to worse MaxEnt
fits. When the definite cluster structure disappears, $\cDt$ grows
dramatically. Thus the existence of well-defined spin clusters is
correlated with the pairwise sufficiency.

\begin{figure}[t]
\centering
\includegraphics[height = 1.5in]{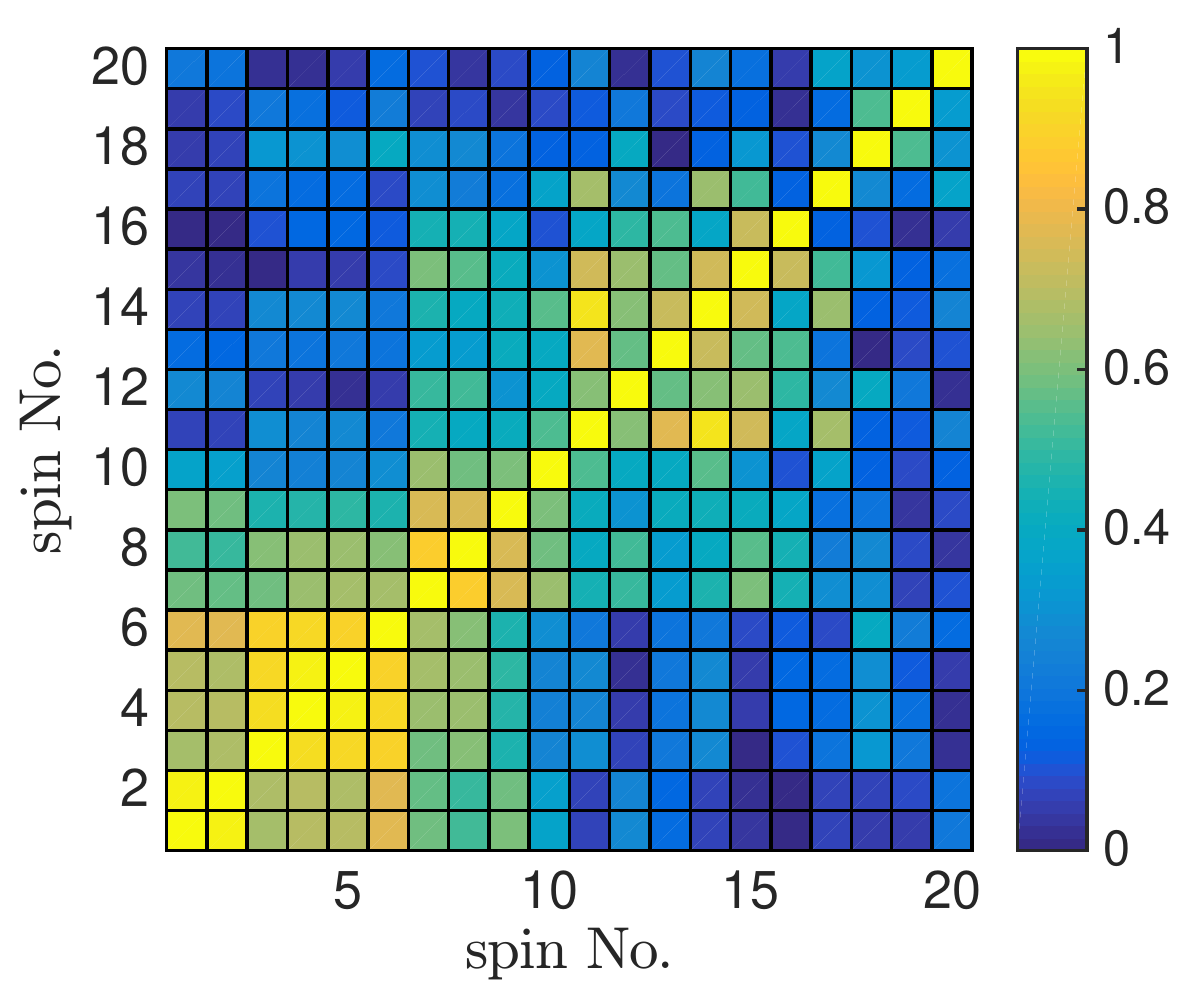}
\caption{\label{correlations3}{\bf A 3-spin network also decouples
    into spin clusters in the pairwise sufficient regime.} For
  brevity, here we show $|c_{ij}|$ only for one network similar to the
  middle panel in Fig.~\ref{correlations}: $S\approx3.5$,
  $\cDo\approx 0.58$, and $\cDt\approx0.038$.  Four to seven partially
  overlapping clusters can be seen. }
\end{figure}

For 3-spin networks, in addition to the pairwise interactions, there
are also nonzero single spin biases in the MaxEnt fits. Thus the
entropy and correlations among spins are generally smaller for the
same $\cDt$. Nonetheless, as seen in Fig.~\ref{correlations3}, the
(fuzzy) cluster structure for these networks is not that much
different from the 4-spin case.

To further explore the network landscapes, we point out that an
inferred symmetric $J_{ij}$ can be rewritten as
\begin{equation}
J_{ij}=\sum_{\nu=1}^N\lambda^{(\nu)} \xi^{(\nu)}_i \xi^{(\nu)}_j,
\end{equation}
where $\lambda^{(\nu)}$ and $\xi^{(\nu)}_i$ are the eigenvalues and
the eigenvectors, correspondingly. The eigenvalues take both large
positive and large negative values for the pairwise sufficient
networks (cf.~Fig.~\ref{eigenvalues}). The negatives correspond to
wells in the landscape, and the positives correspond to peaks. If the
wells and the peaks were clearly separated, then spin configurations
in the vicinity of the well center, distinct from it by just a single
spin flip, would have similar high probabilities (this is what allows
an eigenvector to act as a broad attractor in the Hopfield network
\cite{Hopfield:1982vm}). The repulsive peaks far away from the wells
would have little effect on $\cDt$ since the probability of states
away from the wells is small in the low temperature regime even
without the peaks. In contrast, were spins to form tight clusters,
flipping a single spin would not be allowed. Peaks would be needed to
decrease the probability of such cluster-breaking states, and thus
positive eigenvalues would affect $\cDt$ strongly.

\begin{figure}[t]
\centering
\includegraphics[height = 1.75in]{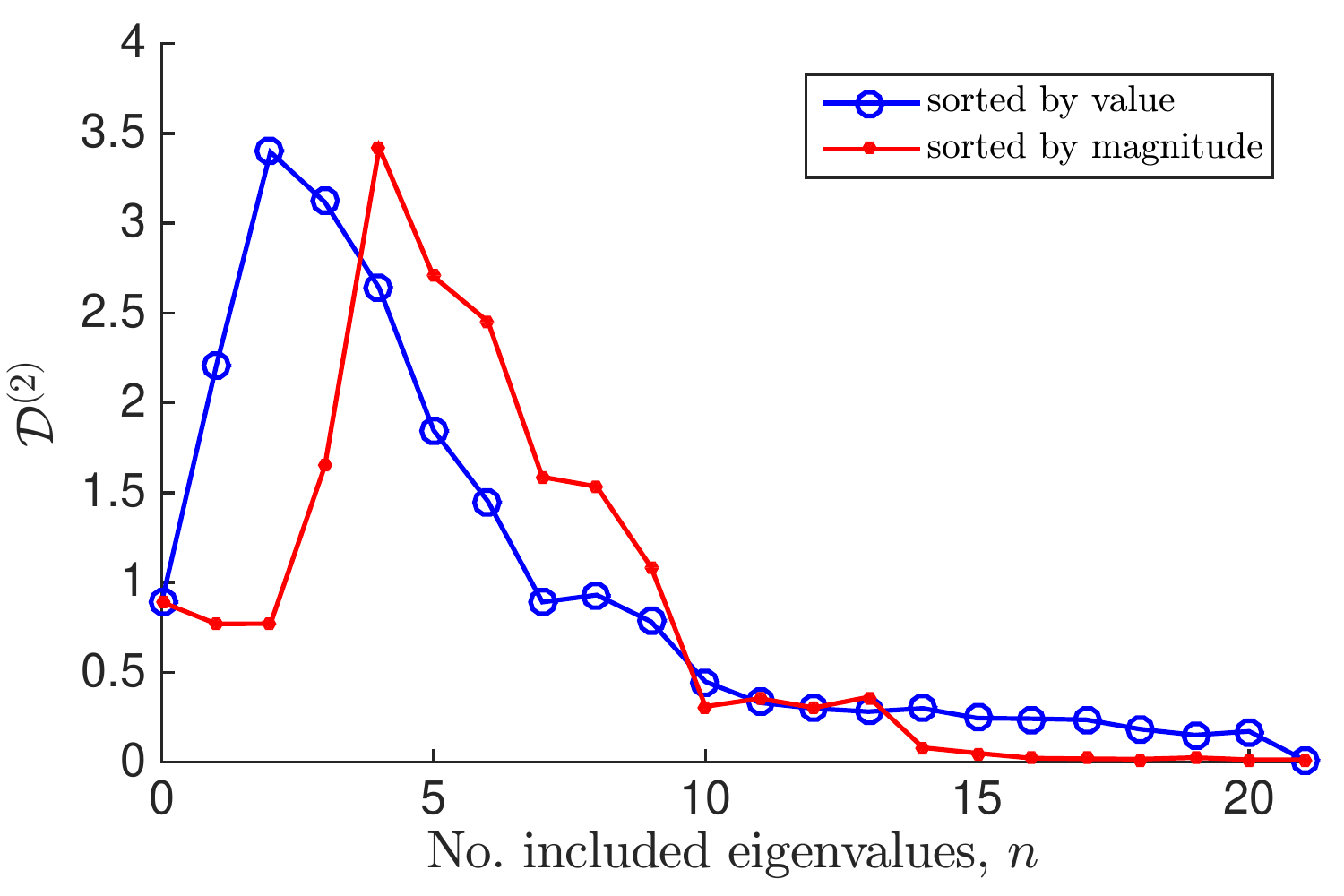}
\caption{\label{D_eigen}{\bf Positive eigenvalues of the MaxEnt
    coupling matrix $J_{ij}$ contribute to the approximation error.}
  We plot the fit error, $\cDt$, as a function of the number of the
  eigenvalues of $J_{ij}$ included in the fit for a 4-spin
  distribution with $N=21$, $\cS\approx 0.1$, and
  $\cDt\approx0.014$. The blue line includes eigenvalues in the order
  from the most negative to the most positive, and the red one
  includes them in the order of their absolute values.  The red line
  reaches the limiting value of $\cDt$ quicker, while the blue one
  requires inclusion of all eigenvalues for this to happen. As
  explained in the text, this is a signature of emergence of spin
  clusters. }
\end{figure}

To verify which of the two scenarios holds, for 4-spin networks, we
construct the coupling matrix and the pairwise MaxEnt distribution
from only $n\le N$ eigenvalues,
\begin{align}
J_{ij}(n)&=\sum_{\nu=1}^{n}\lambda^{(\nu)} \xi^{(\nu)}_i
\xi^{(\nu)}_j,\\
Q^{(2)}(n)&=\frac{1}{Z}\exp\left(-\sum_{i,j}J_{ij}(n)\sigma_i\sigma_j\right).
\end{align}
We then evaluate $\cDt$ between $P$ and $Q^{(2)}(n)$ as a function of
$n$. Figure~\ref{D_eigen} shows this dependence for a typical
pairwise-sufficient distribution and for two different ways of
including eigenvalues into $J$. In the first, we proceed from the most
negative eigenvalue to the most positive one. In the second, we
proceed from the largest magnitude eigenvalue to the smallest
one. Since sorting by magnitude (which includes large positive
eigenvalues earlier) approaches the terminal $\cDt$ faster, wells and
peaks must both affect close spin configurations. This is again
consistent with the clustering picture.

\subsection{The mechanism of emergence of the pairwise sufficiency}
\begin{figure}[t]
\centering
\includegraphics[height = 1.7in]{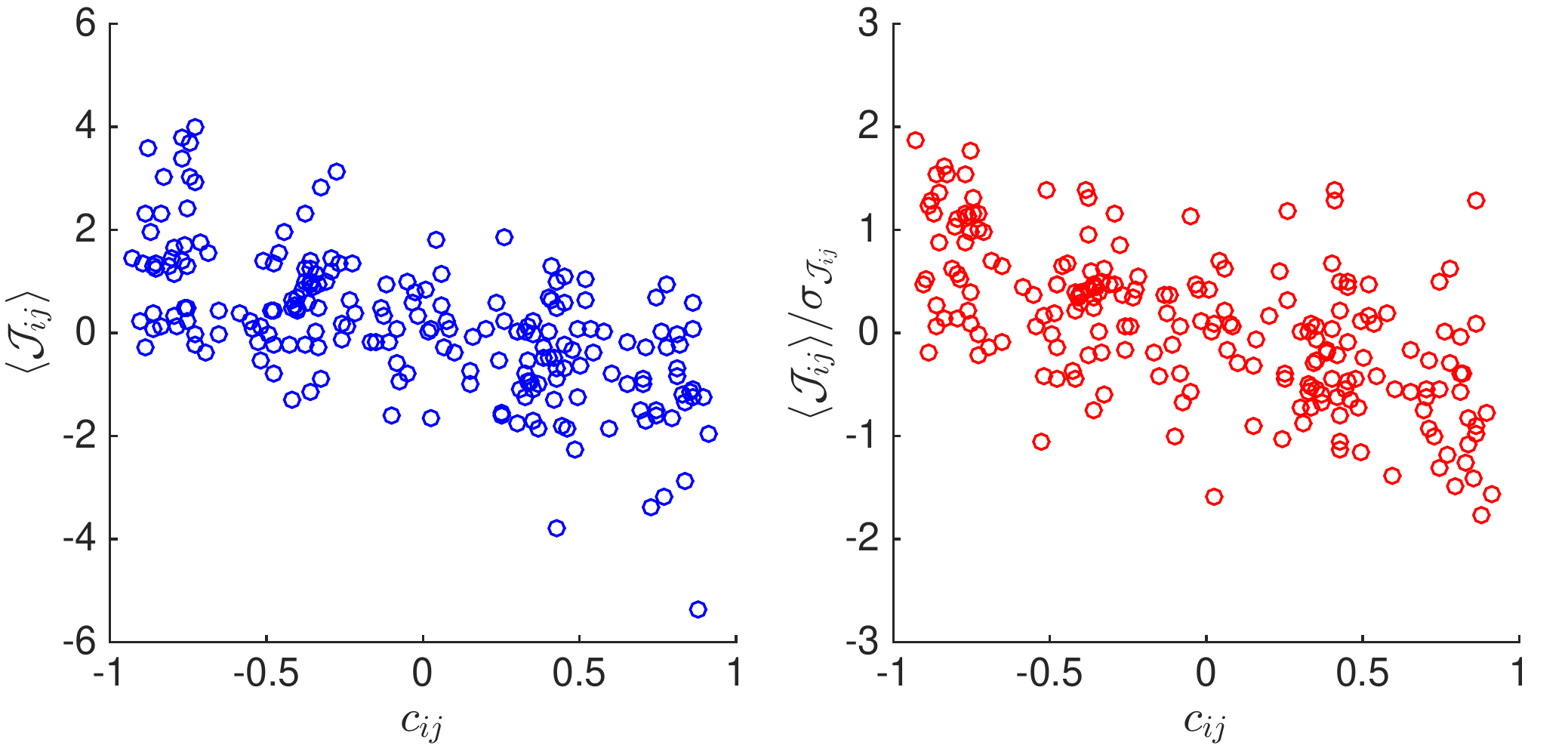}
\caption{\label{cJ}{\bf Formation of effective couplings in $4$-spin
    networks.} For the same network as in Fig.~\ref{correlations},
  middle, we plot $\langle \cJ_{ij}\rangle$ (left) and
  $\langle \cJ_{ij}\rangle$ in units of the standard deviation (right)
  for each pair of spins vs.~$c_{ij}$. For correlated spins, higher
  order coupling terms add up, on average, to strong pairwise
  couplings. Negative $\cJ_{ij}$'s typically correspond to positive
  correlations (and vice versa), as expected.}
\end{figure}

The clustered structure of the network landscapes allows us to propose
a hypothesis for why densely coupled $p$-spin networks exhibit the
pairwise sufficiency. We re-group terms in the energies, which define
$P_3$ and $P_4$ in Eqs.~(\ref{triplet},~\ref{quartic}). For example,
all terms that couple $\sigma_i$ and $\sigma_j$ for the 4-spin network
can be rewritten as
\begin{equation}
  \sigma_i\sigma_j\sum_\mu K_\mu
  \sigma_{\mu_3}\sigma_{\mu_4}\delta_{i,\mu_1}\delta_{j,\mu_2}\equiv\sigma_i\sigma_j\cJ_{ij},
\label{couplings}
\end{equation}
where $\delta_{\cdot,\cdot}$ is a Kronecker delta. (Here we slightly
abused the notation and imposed that $i,j$ only occur in $\mu_1$ and
$\mu_2$.) This equation defines a random coupling $\cJ_{ij}$, which
depends both on the current network state and on the quenched
randomness that went into building the network. For a large number of
couplings, fluctuations in $\cJ_{ij}$ will be large enough so that,
averaged over the accessible network states, $\cJ_{ij}$ stays far from
zero compared to its standard deviation $\sigma_{\cJ_{ij}}$. This
creates large effective pairwise coupling among spins, so that
clusters of spins start behaving coherently. Then the state of every
spin in the cluster can be defined by choosing a cluster
representative, setting its value, and then coupling each cluster
member to the representative through a pairwise interaction --- higher
order couplings are not needed! The pairwise MaxEnt fit is nearly
exact, even though the network is far from frozen since values of the
cluster representatives are not necessarily constrained. We illustrate
this in Fig.~\ref{cJ}, which shows that higher order couplings average
to produce large effective pairwise interactions for correlated spins.

\section{Discussion}

In this numerical study, we showed that pairwise MaxEnt models are
more effective in approximating random $p$-spin networks ($p=3,4$)
than one would naively expect. Even in the worst cases, the error of
such models was rarely above $\cDt\sim0.3$, and it was much lower for
densely coupled networks, with lower entropy per spin. We traced the
emerging pairwise sufficiency to formation of coherent clusters of
spins, largely decoupled among themselves, resulting in a multitude of
dependent attractors for the system. This is not a perturbative effect
and is a new proposal for explaining pairwise sufficiency. Such
collective behavior introduces substantial redundancy, and would allow
error correction. However, this error correction is of a very
different nature compared to, for example, the Hopfield network
\cite{Hopfield:1982vm}.

Does the mechanism presented here explain the pairwise sufficiency in
any {\em real} biological system? This is unclear since our analysis
was limited to specific simulated networks, which may or may not be
good models of real biology. Specifically, the network in the original
paper that observed the pairwise sufficiency \cite{Schneidman:2006he}
had much smaller entropy per spin (neurons rarely fired), and
correlations among spins rarely exceed 0.2. In contrast, while 3-spin
networks in our simulations had smaller $\cS$ and smaller spin-spin
correlations than their 4-spin counterparts, these numbers were still
larger than those in the experiments. At the same time, pairwise
MaxEnt models do not fit experimental data perfectly (certainly worse
than some of our nearly perfect fits) \cite{Tkacik:2014ey}. It may be
that some structural features of real systems allow them to operate at
higher $\cDt$ for smaller $\cS$ compared to the simple models we
investigated here --- and exploring a wider class of networks for
signatures of behaviors that we observed would be the next step. This
is especially important since large coherent deviations from the most
probable state into $10\dots100$ metastable states seems to be a
crucial feature of many experimental systems (such as bursts of neural
activity in the retina that predominantly stays quiet
\cite{Tkacik:2014ey}). Such metastable states far away from the ground
state at least resemble the models that we studied. In addition,
MaxEnt models in other fields may have very different properties
compared to those in neuroscience, including different typical
entropies and correlation strengths. Therefore, we hope that our
models and their generalizations will be able inform interpretation of
experimental data, even if they do not match the experiments in some
important properties.

With (approximate) pairwise sufficiency seen in many collective
biological phenomena, it is important to ask why these systems operate
in the regime that allows it to hold. Indeed, within our model, the
pairwise sufficiency is not generic: low $\cDt$ happens only for small
$\cS$, and preferentially when the strength of the interactions is
high, $\alpha=Ms/N\gg 1$ (cf.~Fig.~\ref{DvsS}). The need for
redundancy and error correction is a potential explanation -- but
there is no obvious reason why the redundancy must result in the
pairwise sufficiency (indeed, simple parity-based codes probably do
not). Taking the improvement in $\cDt$ with the increase in $\alpha$
seriously, we propose a different explanation (a similar argument was
first suggested in Ref.~\cite{lander}). 

One can view evolution as trying to satisfy a growing list of
constraints imposed upon a biological network by its interactions with
the environment. These constraints can include efficient information
processing, low energy consumption, robustness to perturbations,
fitting within a certain physical size, responding quickly enough so
that actions are relevant in the changing world, etc. Some of these
global constraints may be equivalent to a large number of local
constraints. For example, efficient information transmission in the
visual system typically includes removal of redundancy present in the
natural stimuli \cite{Barlow}, which is equivalent to a multitude of
constraints on activities of nearby neurons. When contraints are
added, fewer and fewer states of the network remain
accessible. Importantly, at least for certain abstract constraint
satisfaction problems \cite{Krzakala:2007dd,myers}, before there are
no more states left, the accessible states organize themselves in a
handful of small, well-separated groups. Whether these states are
uncorrelated, or consist of collective flipping of clusters of spins,
they can be well represented by pairwise MaxEnt models. (In the former
case, such MaxEnt model would have a Hopfield network structure
\cite{Hopfield:1982vm}; in the latter, pairwise interactions would
determine cluster assignments.) Therefore, it can be that the pairwise
sufficiency is a signature of a biological network nearing the
unsatisfiability threshold, being pushed towards it by
evolution. Exploring landscapes of satisfiability problems with more
realistic ensembles of constraints (or interactions) and comparing
them to the landscapes observed in experiments would address this
hypothesis.

\begin{acknowledgements}
  We thank Aly Pesic and Daniel Holz, who helped during the early
  stages of this project, and Arthur Lander and Chris Myers, who
  suggested a possible link to evolution. We are grateful to the Emory
  College Emerson Center for Scientific Computing and its funders for
  the help with numerical simulations. The authors were partially
  supported by the James S.\ McDonnell foundation Complex Systems
  award, by the Human Frontiers Science Program, and by the National
  Science Foundation.
\end{acknowledgements}


\end{document}